\documentclass{article}

\usepackage{amsthm}
\usepackage{amsmath}
\usepackage{amssymb}
\usepackage{cite}
\usepackage{url}
\usepackage{graphicx}

\newtheorem{proposition}{Proposition}

\def\code{\mathcal{C}}
\def\egaledef{\triangleq}

\title{Key Reduction of McEliece's Cryptosystem\\ Using List Decoding}
\author{Morgan Barbier\footnote{Computer Science laboratory of \'Ecole
    Polytechnique - LIX - INRIA Saclay - \^Ile de France
    \footnotesize{\tt morgan.barbier@lix.polytechnique.fr}} \and Paulo
  S. L. M. Barreto\footnote{Department of Computer and Digital Systems
    Engineering - Escola Polit\'{e}cnica, University of S\~{a}o Paulo,
    Brazil \footnotesize{\tt pbarreto@larc.usp.br}}}
\date{}

\begin{document}
\sloppypar
\maketitle

\begin{abstract}
  Different variants of the code-based McEliece cryptosystem were proposed to
  reduce the size of the public key. All these variants use very
  structured codes, which open the door to new attacks exploiting
  the underlying structure.
  In this paper, we show that the dyadic variant can be designed to resist
  all known attacks. In light of a new study on list decoding
  algorithms for binary Goppa codes, we explain how to
  increase the security level for given public keysizes.
  Using the state-of-the-art list decoding algorithm instead of unique
  decoding,
  we exhibit a keysize gain of about 4\% for the standard McEliece
  cryptosystem and up to 21\% for the adjusted dyadic variant.
\end{abstract}

\section{Introduction}
\label{Sec:Intro}
The past few years have seen a renewed interest in code-based
cryptosystems due to their resistance to known quantum attacks
\cite{BBD}.
The famous McEliece asymmetric cryptosystem \cite{McE} is
perhaps the most studied of them.
The private key
is the generator matrix of a code $\code$ and the public key is
obtained from this generator matrix by a permutation of its
columns followed by a multiplication by an random invertible
matrix. This
public key is thus a generator matrix of a code $\code '$
equivalent to $\code$. The encryption consists in encoding the plaintext 
into a codeword $c' \in \code'$ using the public key and randomly
adding as many errors as made possible by the decoding algorithm of
$\code$. The
decryption step consists in decoding the cyphertext over $\code$,
thanks to the private key.\\

The McEliece cryptosystem delivers high encryption and
decryption speeds compared to other systems like RSA \cite{MBEprint} but
suffers from the large size of the associated keys which makes it
unpractical. Lately, a lot
of effort has been put into the design of variants based on
different code families in order to reduce the size of the keys.
For example, in 2008, a
solution was proposed for signature schemes using double-circulant
matrices \cite{ACG}. In \cite{BCGO},
the authors proposed a key reduction for the McEliece
cryptosystem using quasi-cyclic alternant codes. The same year, a method
using the sub-family of classical binary Goppa codes, called quasi-dyadic
codes, was introduced in \cite{MBEprint}, adapting the
idea from \cite{BCGO}. Another key reduction technique
formulated in \cite{BergerLoidreau} hides the structure of a subcode
of generalized Reed-Solomon codes.
Generally speaking, all of these key reduction techniques involve the
introduction of some kind of additional structure. As a cryptographic
rule of thumb, the presence of unneeded structure is often seen as
a potential angle of attack. Indeed, cryptanalysts quickly proposed
new structural attacks against the aforementioned variants
\cite{OtmaniTillichDallot,Stern,FOPT}.\\

Roughly speaking, we can distinguish between two types of attacks. The
first type tries
to recover the plaintext from
the cyphertext, without the knowledge of the private key. It is
clear that increasing the number of errors during the encryption step
will make this kind of attacks more difficult. Bernstein, Lange and
Peters contributed to assess the effectiveness of such attacks in
\cite{BLPWCC} by giving asymptotic analysis of different decoding
algorithms for code-based cryptography. Moreover, working within a
strict complexity model, Finiasz and Sendrier exhibited 
lower bounds for system designers \cite{FSLowBound} by taking into
account the costs of the best decoding attacks \cite{Stern}.\\

The second type of attacks consists in retrieving the private key
from the available public one. Such an attack was recently introduced in
\cite{FOPT} and
boils down to computing a Groebner basis to find the structure of an
alternant code. The McEliece variant with the parameters proposed in
\cite{BCGO} is considered to be 
broken by this attack.
While the dyadic instance from \cite{MBEprint} is also vulnerable,
this variant can be made more robust as shown in
Section~\ref{Sec:KR}.\\

This paper is organised as follows. Section~\ref{Sec:LD} is devoted to
the decoding of binary Goppa codes, most precisely on the
correction radius of different decoding
algorithms. In Section~\ref{Sec:KR} we show how the
dyadic variant can be made more secure against \cite{FOPT} and
present our results on keysize
reduction obtained using the best known list decoding algorithm
for the classical and modified, hardened variants of the McEliece
cryptosystem.\\

\section{List decoding of binary Goppa Codes}
\label{Sec:LD}
Since a major part of the cryptanalysis of code-based cryptography is
intimately linked to error correction, a natural idea is to
add as many errors as possible during the encryption step, provided that 
the recipient is still able to correct them. Decoding a random code is
a hard problem; indeed it was shown that decoding general codes is
NP-complete \cite{DecodingNPHard}. 
The McEliece cryptosystem originally used binary Goppa codes. Some
variants are based on different types of codes ({\em e.g.}
\cite{BergerLoidreau}),
but most of them have been broken ({\em e.g.}
\cite{OtmaniTillichDallot}). In the following, we briefly recall
the state of the art of the decoding of binary Goppa codes, which are
perhaps the most promising for McEliece cryptosystems.\\

\begin{figure}[!ht]
  \centering
  \includegraphics[width=9cm]{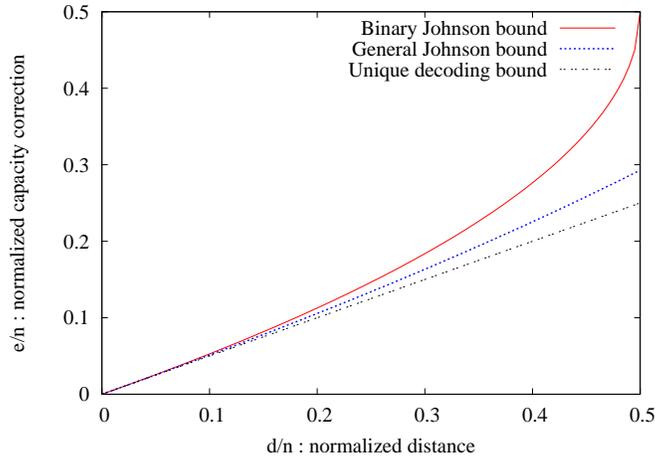}
  \caption{\label{Fig:comparaison}
    Comparison between the unambiguous decoding, generic and binary
    Johnson's bounds.}
\end{figure}

The first algebraic decoding algorithm for classical Goppa codes was
proposed by Patterson in 1975 \cite{Patterson}. This algorithm, 
basically a variation of the Berlekamp-Massey algorithm
\cite{Berlekamp}, runs in quadratic time in the code
length. Patterson's method
performs an unambiguous decoding, up to the error capacity 
$t$ of the code. Since classical Goppa codes are
alternant, that is they are subfield subcodes of generalised Reed-Solomon
codes \cite{Sloane}, we are able to perform the well-known
Guruswami-Sudan list decoding (GS-LD) algorithm \cite{GSLD}. 
This method makes it possible to correct up to the generic Johnson
bound given by $n\left(1 - \sqrt{1- \frac{2t}{n}} \right)$ errors, which is
larger than $t$ (see Figure~\ref{Fig:comparaison}). Consequently, this
type of decoding does not ensure the uniqueness of the returned
codewords anymore. The GS-LD algorithm
is originally not tailored to the binary Goppa codes. Using specific
properties of binary Goppa codes, Bernstein was able to extend
Patterson's algorithm to perform a list decoding up to $n\left(1 -
\sqrt{1- \frac{2t+2}{n}} \right)$ \cite{BernsteinLDGC}, which is
larger than the generic Johnson bound. Recently, a technical report
\cite{ABCLD} revisits previous works to exhibit a list decoding
algorithm for square-free binary Goppa codes which decodes up to the
{\em binary} Johnson bound given by
$\tau_2 \egaledef \frac{n}{2}\left(1 - \sqrt{1- \frac{4t+2}{n}}
\right)$, which is
larger than the two former bounds. As shown in
Figure~\ref{Fig:comparaison}, the closer the normalized distance is to
$0.5$, the better the binary Johnson bound is compared to the others.
We will show in Section~\ref{Sec:KR} that using binary Goppa codes
with normalized minimum distances closer to $0.5$ makes it possible to
correct more errors and ultimately, to reduce the size of the keys.\\

List decoding algorithms basically involve two steps. The first stage
finds, by interpolation, a bivariate polynomial connecting the
received word with the support of the code. The second step consists
in finding the roots of this polynomial. 
The cost of the algorithm from \cite{ABCLD} is dominated by the
interpolation step. This algorithm has an overall
complexity of $\mathcal{O}(n^2\epsilon^{-5})$ and corrects up to
$(1-\epsilon)\tau_2$ errors, where $\tau_2$ is the binary
Johnson bound. Decoding $\tau_2$ errors is obviously
prohibitively expensive but trade-offs between running time and number
of corrected errors are easily achieved, making it possible to keep
the cost of list decoding under control.\\

The classical McEliece or equivalently the Niederreiter cryptosystems
\cite{Nied,equivalenceMcEN} suffer from chosen cyphertext attacks
\cite{CCAttack}. Indeed, since a given plaintext can be encrypted to give
different cyphertexts, an attacker could compare these different
cyphertexts to extract the original plaintext. Different methods were
proposed to make these cryptosystems more robust to chosen cyphertext
attacks \cite{CCA2McE1,CCA2McE2,CCA2McE3} leading to so-called CCA2-secure
variants. 
When adding more errors than can be uniquely corrected, the decryption
step will return a list of potential plaintexts. As already remarked
in \cite{BLP}, CCA2-secure variants make it possible to
distinguish the original plaintext between all candidates returned by
the list decoding algorithm used in the decryption process.
Consequently, it is possible to make the task for an attacker much
more difficult by adding more errors than the correction
capacity. Using CCA2-secure variants and state-of-the-art list
decoding algorithm, these extra-errors only add a small burden on the
recipient to find the original plaintext.\\

\section{Key reduction}
\label{Sec:KR}

Encrypting and decrypting with the McEliece cryptosystem is
significantly faster than with more widespread cryptosystems based on
number theory such as the ubiquitous RSA \cite{MBEprint}.
The main and perhaps only handicap holding back the McEliece
cryptosystem is the substantially larger size of the public keys.
We propose to address this problem not by using a well structured
code as is often the case, but by
adding as many errors as permitted by the best known list decoding
algorithm \cite{ABCLD}. For
a given keysize, this increases the security level. Symmetrically, this
makes it possible to use shorter keys while keeping a similar security
level.
Using a list decoding algorithm can thus lead to
shorter keys at the expense of a moderately increased decryption
time.\\

We focus on the family of square-free binary Goppa codes,
which includes the traditionally used family of irreducible binary
Goppa codes. In this case the error capacity $t$ is equal to $r$ the
degree of Goppa polynomial. The
algorithm decoding the largest number of errors for these codes is
studied in
\cite{ABCLD}. This list decoding algorithm works for all
alternant codes, but using proposition~\ref{Prop:egalite}
improves the correction radius and leads to even shorter keys.
We numerically searched for codes parameters yielding short
keys and correcting up to $\lceil \tau_2 \rceil - 1$.
We illustrate the benefits of list decoding by presenting examples for
both the generic and dyadic variants.

\subsection{Generic variant}
\label{Sub:G}

Tables \ref{Tab:gen}, \ref{Tab:DF} and \ref{Tab:DF16} show the keysize
reduction obtained using
the best known list decoding algorithm \cite{ABCLD}, for 
workfactors (WF) equal to $2^{80}$, $2^{112}$, $2^{192}$ and
$2^{256}$. For each workfactors, McEliece keysizes are given for
Unambiguous Decoding (U.D.) and List Decoding (L.D.). The involved
codes are defined by $m$, the degree of the extension where $G$ and
$\mathcal{L}$ are defined, the length $n$, the dimension $k$,
the degree $r$ of the Goppa polynomial $G$, and $\tau_2$ is the binary
Johnson bound reached by the list decoding algorithm. The workfactors
have been estimated using the complexity model and the lower bounds
given in \cite{FSLowBound}.\\

\begin{table}[h!]
  \begin{center}
    \caption{\label{Tab:gen}
      Comparison between the public keysize of generic McEliece
      cryptosystem using unambiguous and list decoding for given
      workfactors.
    }
    \begin{tabular}{|c|c||c|c|c|c||c|c||c|}
\hline
Method & $m$ & $n$  & $k$  & $r$ & $\tau_2$ & WF      & Keysize & gain (\%)\\
\hline
\hline
U.D.   & 11  & 1893 & 1431 &  42 &          &  80.025 &  661122 & \\
\hline 
L.D.   & 11  & 1876 & 1436 &  40 & 41       &  80.043 &  631840 & 4.43\\
\hline 
\hline 
U.D.   & 12  & 2887 & 2191 &  58 &          & 112.002 & 1524936 & \\
\hline 
L.D.   & 12  & 2868 & 2196 &  58 & 59       & 112.026 & 1475712 & 3.23\\
\hline
\hline
U.D.   & 12  & 3307 & 2515 &  66 &          & 128.007 & 1991880 & \\
\hline 
L.D.   & 12  & 3262 & 2482 &  65 & 66       & 128.021 & 1935960 & 2.81\\
\hline 
\hline 
U.D.   & 13  & 5397 & 4136 &  97 &          & 192.003 & 5215496 & \\
\hline 
L.D.   & 13  & 5269 & 4021 &  96 & 98       & 192.052 & 5018208 & 3.78\\
\hline 
\hline 
U.D.   & 13  & 7150 & 5447 & 131 &          & 256.002 & 9276241 & \\
\hline
L.D.   & 13  & 7008 & 5318 & 130 & 133      & 257.471 & 8987420 & 3.11\\
\hline 
\end{tabular}


  \end{center}
\end{table}

Table~\ref{Tab:gen} refers to the generic McEliece system where the
size of the public keys is given by $(n-k) \times k = mkr$.
As shown in figure~\ref{Fig:comparaison}, using a list decoding
algorithm is all the more interesting as the normalized minimum
distance \mbox{$(2r+1)/n$} gets closer to $0.5$, which has apparently an
adverse effect on the keysize. However, even in this unfavorable case,
we were still able to exhibit a keysize reduction of about $4\%$.

\subsection{Dyadic case}
\label{Sub:DF}

The attack proposed by Faug\`ere, Otmani, Perret and Tillich in
\cite{FOPT} uses Groebner basis computations to
recover the private key from the only knowledge on the public one. 
It was specifically designed to break the compact key McEliece variants 
proposed in \cite{BCGO,MBEprint}, which use the structure of alternant
codes. The variant proposed in \cite{MBEprint} uses binary Goppa codes in dyadic
form, which are also alternant codes. The attack in \cite{FOPT} thus
applies and can recover an equivalent private key in an alternant code
form. However, this is not sufficient to break the system when using
Goppa codes. Indeed, the attack does not directly retrieve the Goppa
polynomial $G$ of degree $r$ which is crucial to decode
\cite{Patterson,ABCLD},  but finds a generator matrix of
an alternant code without a Goppa structure and with designed minimum
distance $r+1$. However, when using a
Goppa code, the private key is a generator matrix of a code with
designed minimum distance $2r+1$ thanks to the following proposition,
demonstrated in \cite{ABCLD,Wild}:

\begin{proposition}
  \label{Prop:egalite}
  Let $G$ be a square-free polynomial in $\mathbb{F}_{2^m}$ and
  $\mathcal{L}$ be a list of $n$ elements of $\mathbb{F}_{2^m}$ which are
  not roots of $G$. Then
  $$
  \Gamma(\mathcal{L},G) = \Gamma(\mathcal{L},G^2),
  $$
  where $\Gamma(\mathcal{L},G)$ is the Goppa code generated by
  $\mathcal{L}$ and $G$.
\end{proposition}

The direct consequence is that the attacker won't be able to
decode. Indeed, this attack retrieves $n/r$
variables $Y$ and $n$ variables $X$ such that $Y_i = G(X_i)^{-1}$. In
order to protect against a potential interpolation of  the
Goppa polynomial $G$ of degree $r$, we impose that $r+1 >
\frac{n}{r}$, that is $r(r+1)>n$. Consequently, this attack does not
totally break the McEliece variant based on dyadic forms.
Moreover, as stated in \cite{FOPT}, the attack becomes unpractical,
for the moment,
when the extension degree $m$ is greater than $16$. Working with such an
extension degree slightly increases
the public keysize of McEliece compared to the parameters proposed in
\cite{MBEprint}, while staying drastically smaller than with the generic
form, as shown in tables \ref{Tab:gen}, \ref{Tab:DF} and
\ref{Tab:DF16}.\\

\begin{table}[h!]
  \begin{center}
    \caption{\label{Tab:DF}
      Comparison between the public keysize of dyadic McEliece
      cryptosystem with $r(r+1) > n$ using unambiguous and list
      decoding for given workfactors.
    }
    
\begin{tabular}{|c|c||c|c|c|c||c|c||c|}
\hline
Method & $m$ & $n$  & $k$  & $r$ & $\tau_2$ & WF      & Keysize & gain (\%)\\
\hline
\hline
U.D. & 11 & 1792 & 1088 & 64 &  & 82.518 & 11968 & \\
\hline 
L.D. & 11 & 1728 & 1024 & 64 & 67 & 82.976 & 11264 & 5.88\\
\hline 
\hline 
U.D. & 12 & 2944 & 1408 & 128 & & 116.735 & 16896 & \\
\hline 
L.D. & 13 & 2816 & 1280 & 128 & 134 & 113.896 & 15360 & 9.09\\
\hline 
L.D. & 13 & 7680 & 1024 & 512 & 552 & 113.084 & 13312 & 21.21\\
\hline 
\hline 
U.D. & 12 & 3200 & 1664 & 128 & & 131.235 & 19968 & \\
\hline 
L.D. & 12 & 3072 & 1536 & 128 & 134 & 129.745 & 18432 & 7.69\\
\hline 
\hline 
U.D. & 13 & 5888 & 2560 & 256 &  & 205.804 & 33280 & \\
\hline 
L.D. & 13 & 5632 & 2304 & 256 & 269 & 199.473 & 29952 & 10.00\\
\hline 
\hline 
U.D. & 15 & 11264 & 3584 & 512 & & 279.002 & 53760 & \\
\hline 
L.D. & 15 & 10752 & 3072 & 512 & 539 & 258.223 & 46080 & 14.29\\
\hline 
\end{tabular}

  \end{center}
\end{table}

\begin{table}[h!]
  \begin{center}
    \caption{\label{Tab:DF16}
      Comparison between the public keysize of dyadic McEliece
      cryptosystem with $m \ge 16$ using unambiguous
      and list decoding for given workfactors.
    }
    \begin{tabular}{|c|c||c|c|c|c||c|c||c|}
\hline
Method & $m$ & $n$  & $k$  &  $r$ & $\tau_2$ & WF      & Keysize & gain (\%)\\
\hline
\hline
U.D.   & 16  & 5120 & 1024 &  256 &          &  81.765 & 16384 & \\
\hline 
L.D.   & 16  & 5120 & 1024 &  256 & 134      &  86.216 & 16384 & 0\\
\hline
\hline
U.D.   & 16  & 3840 & 1792 &  128 &          & 113.785 & 28672 & \\
\hline 
L.D.   & 16  & 5632 & 1536 &  256 & 269      & 116.400 & 24576 & 14.29\\
\hline 
\hline 
U.D.   & 16  & 5888 & 1792 &  256 &          & 132.470 & 28672 & \\
\hline 
L.D.   & 16  & 9728 & 1536 &  512 & 542      & 133.534 & 24576 & 14.29\\
\hline
\hline
U.D.   & 16 & 10752 & 2560 &  512 &          & 199.067 & 40960 & \\
\hline 
L.D.   & 16 & 10752 & 2560 &  512 & 539      & 209.414 & 40960 & 0\\
\hline
\hline
U.D.  & 16  & 11776 & 3584 &  512 &          & 264.846 & 57344 & \\
\hline 
L.D.  & 16  & 19456 & 3072 & 1024 & 1085     & 267.203 & 49152 & 14.29\\
\hline 
\end{tabular}


  \end{center}
\end{table}

In table~\ref{Tab:DF}, we look at the case of the dyadic variant with
countermeasure $r(r+1) > n$. The size of the public key now becomes 
$mk$ \cite{MBEprint}, removing the conflicting constraints on $r$.
Key reduction up to $21\%$ can now be achieved. Finally, results on
the dyadic variant
with countermeasure $m \ge 16$ are presented in
table~\ref{Tab:DF16}. As previously discussed, we expect better
reductions than in the generic case. Indeed, our experiments showed
a key reduction of more than $14\%$. Note that in this case, the degree
$r$ of the Goppa polynomial is the same as the dimension $k$ of the
code. This is easily explained:
the large extension degree becomes such a strong constraint on the
parameters that it removes all freedom when choosing the code
dimension.\\

Table~\ref{Tab:LogD} displays the recommended keysizes for cryptosystems
based on the Discrete Logarithm Problem over finite fields (DLP), for
different security
levels \cite{NIST,RFC}. For the sake of comparison, we also include
the smallest keysizes obtained with McEliece variants although 
in all impartiality,
 it should be stressed that we lack sufficient perspective to
correctly assess the true security level of these fairly new variants.
While the keysizes for the McEliece cryptosystem are
still larger than their discrete logarithm
counterparts, the gap significantly narrows when going at higher
security levels. Moreover, the costs for McEliece encryption and
decryption rise much more slowly with the security level than they
do with DLP based or RSA systems \cite{MBEprint}.

\begin{table}[!ht]
  \centering
  \caption{\label{Tab:LogD}
    Keysize comparison between cryptosystem based on discrete
    logarithm over finite fields and McEliece cryptosystem using list
    decoding.
  }
  \begin{tabular}{|c||c|c||c|}
    \hline
    Security level & Discrete Logarithm & McEliece & ratio\\
    \hline
    \hline
     80 &  1024 & 11264 & 11.0\\
    \hline
    112 &  2048 & 13312 & 6.5\\
    \hline
    128 &  3072 & 18432 & 6.0\\
    \hline
    192 &  7680 & 29952 & 3.9\\
    \hline
    256 & 15360 & 46080 & 3.0\\
    \hline
  \end{tabular}
\end{table}

\section{Conclusion}
In light of the recent study on the list decoding of
binary Goppa codes \cite{ABCLD}, we compared the size of public keys for
different variants of the McEliece cryptosystem. We showed that using list
decodable codes in McEliece cryptosystems deliver compelling benefits.
We explained how to secure
the dyadic variant against currently known attacks while 
reducing the size of the keys using list decoding. For example, for a
workfactor of $2^{80}$, list decoding lowers the
public keysize from 661,122 bits for the generic
variant to  11,264 bits for the dyadic variant. It is worth mentioning
that contrary to previous attempts at reducing the
McEliece keysizes, using list decoding does not introduce any
additional structure that could be used to attack the system.\\

\section*{Acknowledgements}
The authors would like to thank Matthieu Finiasz, J\'er\^ome Milan,
Rafael Misoczki and Ayoub Otmani for stimulating discussions and improving the
editorial quality, and to express his gratitude to Nicolas Sendrier who kindly
let us build on his software.

The second author (P.~Barreto) is supported by the Brazilian National Council
for Scientific and Technological Development (CNPq) under research productivity
grant 303163/2009-7.

\bibliographystyle{plain}
\bibliography{biblio}

\end{document}